\documentclass[12pt]{article}
 \pdfoutput=1

\usepackage{amsmath,amssymb,amsfonts}
\usepackage{graphicx}
\usepackage{color}

\newcommand{\cref}[1]{Chapter~\ref{c.#1}}

\newcommand{\barray}{\begin{eqnarray}}
\newcommand{\earray}{\end{eqnarray}}
\newcommand{\nn}{\nonumber \\}

\newcommand{\beq}{\begin{equation}}
\newcommand{\eeq}{\end{equation}}
\newcommand{\ba}{\begin{array}}
\newcommand{\ea}{\end{array}}
\newcommand{\bea}{\begin{eqnarray}}
\newcommand{\eea}{\end{eqnarray} }
\newcommand{\be}{\begin{eqnarray}}
\newcommand{\ee}{\end{eqnarray} }
\newcommand{\bal}{\begin{align}}
\newcommand{\eal}{\end{align}}
\newcommand{\bi}{\begin{itemize}}
\newcommand{\ei}{\end{itemize}}
\newcommand{\ben}{\begin{enumerate}}
\newcommand{\een}{\end{enumerate}}
\newcommand{\bc}{\begin{center}}
\newcommand{\ec}{\end{center}}
\newcommand{\bt}{\begin{table}}
\newcommand{\et}{\end{table}}
\newcommand{\btb}{\begin{tabular}}
\newcommand{\etb}{\end{tabular}}

\def\cl{{\mathcal L}}

\def\mev{\, {\rm MeV}}
\def\gev{\, {\rm GeV}}

\newcommand\simlt{\stackrel{<}{{}_\sim}}
\newcommand\simgt{\stackrel{>}{{}_\sim}}

\newcommand{\eps}{\epsilon}

\numberwithin{equation}{section}

\begin{document}
\begin{titlepage}
\vspace{-1cm}
\begin{flushright}
\small
\end{flushright}
\vspace{0.2cm}
\begin{center}
{\Large \bf Interpreting the Higgs}
\vspace*{0.2cm}
\end{center}
\vskip0.2cm

\begin{center}
{\bf  Dean Carmi$^{a}$,  Adam Falkowski$^{b}$, Eric Kuflik$^{a}$, and Tomer Volansky$^{a}$}

\end{center}
\vskip 8pt

\begin{center}
{\it $^{a}$ Raymond and Beverly Sackler School of Physics and Astronomy, Tel-Aviv University, Tel-Aviv 69978, Israel}  \\
{\it $^{b}$ Laboratoire de Physique Th\'eorique d'Orsay, UMR8627--CNRS,\\ Universit\'e Paris--Sud, Orsay, France}
\end{center}

\vspace*{0.3cm}

\vglue 0.3truecm

\begin{abstract}
The LHC  and Tevatron Higgs data are interpreted as  constraints on an effective theory of a Higgs boson with mass  $m_h \simeq 125$ GeV. 
We  focus on the   $h \to \gamma \gamma$, $h\to ZZ^* \to 4l$,  and  $h\to WW^* \to 2 l 2\nu$ channels at the LHC, and the $b \bar b$ channel at the Tevatron, which are currently  the most sensitive probes of a Higgs with such a mass.  
Combining the available data in these channels, we derive the preferred regions of the parameter space of the effective theory.   
We further provide the mapping between the effective theory and the relevant Higgs event rates, facilitating future extraction of the preferred region by the ATLAS and CMS collaborations.

\vspace{0.1cm}

{\footnotesize \it Contribution to the proceedings of the XXVI Rencontres de Physique de la Vall\'ee d'Aoste, La Thuile 2012. } 
\end{abstract}

\end{titlepage}

\section{Introduction} 

Discovering the Higgs boson Êand measuring its properties is currently the key objective of the high-energy physics program. 
Within the Standard Model (SM), the coupling to the Higgs boson is completely fixed by the particle mass.
This is no longer the case in many scenarios beyond the SM, where the Higgs couplings to the SM gauge bosons and fermions may display sizable departures from the SM predictions.
Indeed, precision studies of the Higgs couplings may be the shortest route to new physics.

Recently, ATLAS \cite{ATLAS_combo} and CMS  \cite{CMS_combo} have reported the results of Higgs searches based on  5~fb$^{-1}$ of LHC data 
while CDF  and D0 presented  Higgs searches based on 10~fb$^{-1}$ of Tevatron data \cite{Tev_combo}.   
The results suggest the existence of a Higgs boson with $m_h \approx 125$ GeV manifesting itself in the diphoton and 4-lepton final states at the LHC, and in the $b\bar b$ final state at the Tevatron.  
Assuming these signals are indeed due to a  Higgs boson, it is natural to ask the following questions: 
\bi
\item Are the experimental data consistent with the predictions of the SM Higgs?  
\item Do the data favor or disfavor any particular scenarios beyond the SM?
\item What are the implications of the Higgs data for new physics models addressing the naturalness problem of  the SM?  
\ei 
A convenient framework to address these questions is that of an effective theory describing general interactions of a light Higgs boson with matter.  
In this approach, laid out in Section 2, a small number of couplings, $c_i$, captures the leading-order Higgs interactions relevant for the current LHC and Tevatron searches. 
We explain how to express  the event rates in various Higgs search channels, which are directly observable in colliders, in terms of the parameters $c_i$. 
Given the event rates measured by experiments and the corresponding errors (assumed to be Gaussian), we can construct the likelihood functions in the space spanned by $c_i$. 
Since the number of observables is larger than the number of parameters $c_i$, with enough data we will be able to fit all the leading-order parameters of the effective theory.   

To illustrate this approach,  in Section 3  we construct the likelihood function using the LHC and Tevatron Higgs results in 5 search channels that are currently most sensitive to the signal of a 125 GeV Higgs: 
\bi 
\item  A combination of the inclusive diphoton channels in ATLAS  \cite{ATLAS_gaga} and CMS \cite{CMS_gaga}. 
\item  The dijet tag exclusive diphoton channel in CMS \cite{CMS_gaga}. 
\item  A combination of the inclusive $ZZ \to 4l$ channels in ATLAS   \cite{ATLAS_zz}  and CMS  \cite{CMS_zz}.
\item A combination of the inclusive $WW \to 2l 2\nu$ channel in ATLAS \cite{ATLAS_ww} and CMS  \cite{CMS_ww} . 
\item  The W/Z associated Higgs production in the $b \bar b$ channel at the Tevatron  \cite{Tev_combo}.
\ei 
The likelihood function we construct can be used to identify the best-fit regions of the effective theory (for the time-being constrained to a 4-dimensional subspace, until more Higgs data become available). 
In addition to the  channels discussed here, one may consider other available Higgs measurements  (e.g. the $b\bar b$ and $\tau^+ \tau^-$ channel at the LHC, the $W^+W^-$ and the diphoton channel at ther Tevatron, etc.).  
Those,  however, are currently less sensitive to a 125 GeV Higgs, and including them does not alter the fits significantly.   

The main goal of this note is to collect the formulae needed in order to map the Higgs effective theory to rates measured at colliders, 
in the aim of of helping the experimental collaborations to present a similar but more refined analysis in the near future. 
The formalism we propose in Section~2 is  based on ref. \cite{Carmi:2012yp}, while the fits presented in Section~3 are updated with the Higgs  search results that subsequently appeared  in refs.  \cite{Tev_combo,ATLAS_ww}. 
For other, partly overlapping theoretical analyses of  the 125 GeV Higgs-like excess,  see refs. \cite{Azatov:2012bz,Giardino:2012ww,Rauch:2012wa}.

\section{Formalism} 

We first lay out in some detail the formalism.  
We describe interactions of the Higgs boson with matter using an effective theory approach where a small number of leading order operators captures the salient features of Higgs phenomenology. 
Given the effective action, we derive the relevant production and decay rates as a function of the effective theory couplings. 
With these relations at hand, one can then construct the coupling-dependent likelihood function for a set of measurements,  allowing for bounds to be placed on these couplings and the best-fit regions to be identified. {It is worth stressing that the effective couplings  (and not only the ratios of couplings, as is sometimes believed)  can be constrained by experiment, even if the total Higgs width cannot be directly measured.}  

\subsection{Lagrangian} 

We introduce the effective Lagrangian  defined at the scale of $\mu=m_h$, 
\barray
\label{eq:1}
\cl_{eff}  &=  &
c_V  {2 m_W^2  \over v}  h  \,   W_\mu^+ W_\mu^-  +  c_V   {m_Z^2 \over v} h  \,  Z_\mu  Z_\mu  -  c_{b}  {m_b \over v } h \,\bar b b    -  c_{\tau}  {m_\tau \over v } h  \,  \bar \tau \tau  -  c_{c}  {m_c \over v } h \,   \bar c c   
  \\  \nonumber &&
 + c_{g}  {\alpha_s \over 12 \pi v} h \, G_{\mu \nu}^a G_{\mu \nu}^a  +  c_{\gamma} { \alpha \over \pi v} h \, A_{\mu \nu} A_{\mu \nu}\, 
-  c_{inv}   h \,   \bar \chi \chi  \ .
\earray
This Lagrangian describes the interactions of a light Higgs scalar with matter,  providing a very general and convenient framework for interpreting the current Higgs searches at the LHC and Tevatron.\footnote{A tacit assumption is we use the effective Lagrangian to study processes where the Higgs boson is dominantly produced near threshold. 
For exclusive processes requiring Higgs produced with a very larger  boost, $p_{T,h} \gg m_h$, the contibution of higher order operators may be quanitatively important. }
The couplings of the Higgs boson are allowed to take arbitrary values, parametrized by $c_i$. 
To be even more general, we also allow for a coupling  to weakly interacting stable particles $\chi$, leading to an invisible Higgs partial width.   
This effective approach harbors very few theoretical assumptions. 
One is that the Higgs interactions with $W$ and $Z$ obey custodial symmetry which,  assuming $h$ is a singlet of the custodial $SU(2)$, implies  $c_W=c_Z \equiv c_V$. 
Relaxing that condition would lead to a quadratically divergent 1-loop contribution to the T-parameter, leading to a tension with electroweak precision measurements (see however ref.~\cite{Farina:2012ea}). 
Another theoretical assumption is that the Higgs width is dominated by decays into up to 2 SM particles; more sophisticated BSM scenarios may predict cascade decays into multiple SM particles which would require a separate treatment.  
Finally, we assummed that the Higgs is a positive-parity scalar; more generally, one could allow for pseudoscalar interactions. 

The top quark has been integrated out in Eq.~\eqref{eq:1}  (assuming $m_h < 2 m_t$) and its effects are included in the effective dimension-5 Higgs couplings parametrized by $c_g$ and $c_\gamma$.
However these 2 couplings may well receive additional contributions from integrating out new physics, and therefore are also kept as free parameters. 
At the same order one could include the dimension-5 Higgs coupling  to $WW$ and $ZZ$, however their effects can be in most cases neglected in comparison with the contribution proportional to $c_V$.  We therefore  omit them for simplicity. The Lagrangian should be extended by the  dimension-5 coupling to $Z \gamma$, once measurements in this channel become available. Obviously, to describe   the $t \bar t$ associated Higgs production  process, which may be observable in the 14 TeV LHC run, one would have to integrate the top quark back in.  

In the SM, the terms in the first line of eq.~\ref{eq:1} arise at tree-level: 
$c_{V,\rm SM}  =  c_{b,\rm SM} =  c_{\tau,\rm SM} =  c_{c,\rm SM}=1$. 
The following 2 terms arise at 1 loop and are dominated by the contribution of the top quark:  
$c_{g,\rm SM} \simeq 1.03$, $c_{\gamma,\rm SM}  \simeq 0.23$.
Finally, $ c_{inv, \rm SM} =0$.\footnote{But note that even in the SM there is a small invisible width via the tree-level $h \to ZZ^* \to 4\nu$ and the 1-loop $h \to 2\nu$ decay modes.}


\subsection{Decay} 

With the help of  the effective theory parameters, $c_i$, we can easily write down the partial Higgs decay widths relative to the SM value. 
Starting with the decays mediated by the lower-dimenensional interactions in the first line of Eq.~\eqref{eq:1} we have,
\beq
\label{eq:h45}
\Gamma_{bb} \simeq |c_b|^2\Gamma_{bb}^{\rm SM} ,
\ \ \Gamma_{\tau \tau}  \simeq  |c_\tau|^2\Gamma_{\tau \tau}^{\rm SM}, 
  \ \ \Gamma_{WW}=|c_V|^2\Gamma_{WW}^{\rm SM} ,\ \ 
  \Gamma_{ZZ}=|c_V|^2\Gamma_{ZZ}^{\rm SM},
\eeq
where the SM widths for $m_h=125$ GeV, are given by \cite{Dittmaier:2011ti}
\beq
\label{eq:h79}
\Gamma_{bb}^{\rm SM}=2.3 \mev,\ \  \Gamma_{\tau \tau}^{\rm SM}=0.25\mev,
\ \ \Gamma_{WW}^{\rm SM}=0.86 \mev , \ \ \Gamma_{ZZ}^{\rm SM}=0.1\mev. 
\eeq 
Strictly speaking, Eq.~\eqref{eq:h45}  is valid at leading order. 
However, higher order diagrams which involve one $c_i$ insertion leave these relations intact. 
Thus, Eq.~\eqref{eq:h45}  remains true when higher order QCD corrections are included.   
The decays to gluons and photons are slightly more complicated because, apart from the dimension-5 effective coupling proportional to $c_g, c_\gamma$, they receive contribution from the loop of the particles present in Eq.~\eqref{eq:1}.  One finds 
\bea
\label{eq:ggrate}
\Gamma_{gg}={|\hat c_g|^2 \over|\hat c_{g,\rm SM}|^2 } \Gamma_{g g}^{\rm SM}, 
\qquad  
\Gamma_{\gamma \gamma}={|\hat c_\gamma|^2 \over|\hat c_{\gamma,\rm SM}|^2 } \Gamma_{\gamma \gamma}^{\rm SM}, 
\eea 
where, keeping the leading 1-loop contribution in each case one finds,  
\bea
\hat{c}_{g} &=&   c_g + c_b A_f(\tau_b) + c_c A_f(\tau_c), 
\\ 
\hat{c}_{\gamma} &=&   c_\gamma  + c_V A_v(\tau_W) + \frac{1}{18}c_b A_f(\tau_b) + \frac{2}{9}c_c A_f(\tau_c) + \frac{1}{6} c_\tau A_f(\tau_\tau). 
\eea
Above we introduced the customary functions describing the 1-loop contribution of fermion and vector particles to the triangle decay diagram, 
\bea 
A_f(\tau) &\equiv& \frac{3}{2\tau^2} \left [  (\tau-1)f(\tau)  + \tau \right ], 
\nonumber \\ 
A_v(\tau) &\equiv& \frac{-1}{8\tau^2}\left[3(2\tau-1)f(\tau)+3\tau+2\tau^2\right], 
\nonumber \\ 
f(\tau)
&\equiv&  \left\{ \begin{array}{lll}
{\rm arcsin}^2\sqrt{\tau} && \tau \le 1 \\ -\frac{1}{4}\left[\log\frac{1+\sqrt{1-\tau^{-1}}}{1-\sqrt{1-\tau^{-1}}}-i\pi\right]^2 && \tau > 1 \end{array}\right. , 
\eea
and $\tau_i  = m_h^2/4m_i^2$. 
Numerically, for $m_h \simeq 125 \gev$,  $A_v(\tau_W) \simeq  -1.04$, $A_f(\tau_b) \simeq -0.06 + 0.09 i$. 
so that   $\hat{c}_{g}  \simeq  c_g - 0.06 c_b$ and  $\hat{c}_{\gamma}  \simeq  c_\gamma - c_V$. 
In the SM $c_g$ and $c_\gamma$ arise from integrating out the top quark, 
thus   $c_{g,\rm SM} = A_f(\tau_t)  \approx 1.03$, and  $c_{\gamma,\rm SM} =  (2/9)c_{g,\rm SM} $.
The SM witdhs are $ \Gamma_{gg}^{\rm SM}\simeq 0.34 \mev$ and  $\Gamma_{\gamma \gamma}^{\rm SM}\simeq 0.008 \mev$. 

In order to compute the branching fractions in a given channel we need to divide the corresponding partial width by the total width,
\beq 
{\rm Br}(h \to i \bar i )  \equiv  {\rm Br}_{ii}= \frac{\Gamma_{ii}}{\Gamma_{tot}}\,.
\eeq
The latter includes the sum of the width in the visible channels, and the invisible width which, for $m_h=125$ GeV, is  $\Gamma_{inv}  \simeq 1.2\times 10^3 c_{inv}^2  \Gamma_{tot}^{\rm SM}$.
We can write it as  
\beq
\label{eq:ctotdef}
\Gamma_{tot}=  |C_{tot}|^2 \Gamma_{tot}^{\rm SM}\,,
\eeq
  where, for $m_h=125$ GeV,  $\Gamma_{tot}^{\rm SM}\simeq 4.0 \mev$, and 
\bea
\label{e.ctot}
|C_{tot}|^2 &\simeq& |c_b|^2  {\rm Br}_{bb}^{\rm SM} + |c_V|^2 \left({\rm Br}_{WW}^{\rm SM}+{\rm Br}_{ZZ}^{\rm SM}\right)+ \frac{|\hat c_g|^2}{\left|\hat c_g^{\rm SM}\right|^2} {\rm Br}_{gg}^{\rm SM} +   |c_\tau|^2 {\rm Br}_{\tau\tau}^{\rm SM} +  |c_c|^2 {\rm Br}_{cc}^{\rm SM} +  { \Gamma_{inv} \over   \Gamma_{tot}^{\rm SM} } 
\nonumber \\
&\simeq& 0.58 |c_b|^2+0.24 |c_V|^2 + 0.09 \frac{|\hat c_g|^2}{\left|\hat c_g^{\rm SM}\right|^2} +  0.06  |c_\tau|^2  +  0.03 |c_c|^2 +{ \Gamma_{inv} \over   \Gamma_{tot}^{\rm SM} } . 
\nonumber 
\eea
Typically, the total width is dominated by the decay to b-quarks and $\Gamma_{tot} \sim c_b^2$, however this scaling may not be valid if the Higgs couples more weakly to  bottoms ($c_b \simlt 0.7$), or more strongly to  gauge fields ($c_V \simgt 1.4$), or if it has a significant invisible width ($c_{inv} \simgt 0.03$). 

\subsection{Production} 

Much like the decay rates, one can express the relative cross sections for the Higgs production  processes in terms of the parameters $c_i$. 
For the LHC and the Tevatron the currently relevant partonic processes are 
\bi 
\item Gluon fusion (ggF), $g g \to h $+jets,
\item Vector boson fusion (VBF),  $q q \to h qq$+jets, 
\item Vector boson associate production (VH), $q \bar q \to h V$+jets    
\ei  
The relative cross sections in these channels can be approximated at tree-level by,
\beq
\label{eq:prod}
{\sigma_{ggF} \over \sigma_{ggF}^{\rm SM}} \simeq {|\hat c_g|^2 \over |\hat c_{g,\rm SM}|^2 }
\quad
{\sigma_{VBF} \over \sigma_{VBF}^{\rm SM}} \simeq |c_V|^2, 
\quad 
{\sigma_{VH} \over \sigma_{VH}^{\rm SM}} \simeq |c_V|^2. 
\eeq 
For $m_h = 125 \gev$, the 7 TeV proton-proton cross sections are  $\sigma_{ggF}^{\rm SM} = 15.3$~pb,  $\sigma_{VBF}^{\rm SM} = 1.2$~pb and $\sigma_{VH}^{\rm SM} = 0.9$~pb \cite{Dittmaier:2011ti}. 
Using Eq.~\eqref{eq:prod}, we find the total inclusive $pp \to h$ cross section $\sigma_{tot}$, 
\beq
\label{eq:sigtot}
{\sigma_{tot}  \over \sigma_{tot}^{\rm SM} } \simeq {|\hat c_g|^2  \sigma_{ggF}^{\rm SM} /|\hat c_{g,\rm SM}|^2 + |c_V|^2  \sigma_{VBF}^{\rm SM} + |c_V|^2 \sigma_{VH}^{\rm SM} \over  
 \sigma_{ggF}^{\rm SM} +  \sigma_{VBF}^{\rm SM} +  \sigma_{VH}^{\rm SM}} \,,
\eeq 
is typically dominated by the gluon fusion process, and therefore it scales as $\sigma_{tot} \sim c_g^2$.

\subsection{Rates} 

The event count in experiments depends on the product of the Higgs branching fractions and the production cross section in a given channel. 
Typically, the results are presented as constraints on R defined as the event rates relative to the rate predicted by the SM (sometimes denoted as $\hat \mu$). 
These rates  can be easily expressed in terms of the parameters of our effective Lagrangian in Eq.\eqref{eq:1}. 
First, the ATLAS and CMS searches in the $\gamma \gamma$, $Z Z^*$  and $WW^*$ channels probe, to a good approximation, the inclusive Higgs cross section.    
Thus, we have 
\begin{eqnarray}
\label{eq:4}
R_{VV^*}^{\rm inc} &\equiv& {\sigma_{tot}  \over \sigma_{tot}^{SM} }{ {\rm Br}(h \to VV^*) \over {\rm Br}_{SM}(h \to VV^*) } 
\simeq  \left|\frac{\hat c_g c_V}{\hat c_{g,\rm SM} C_{tot}} \right|^2,    
\nn
R_{\gamma \gamma}^{\rm inc} &\equiv& {\sigma_{tot} \over \sigma_{tot}^{SM} }{ {\rm Br}(h \to \gamma \gamma ) \over  {\rm Br}_{SM}(h \to \gamma \gamma) }
\simeq    \left | {\hat c_g  \hat c_{\gamma} \over  \hat c_{g,SM} \hat c_{\gamma,SM} C_{tot}}\right|^2. 
\eea 
The approximation holds assuming the Higgs production remains dominated by the gluon fusion subprocess.   The more precise relations (which we use in our fits) can be easily extracted by substituting Eqs.~\eqref{eq:h45}, \eqref{eq:ggrate}, \eqref{eq:ctotdef}, \eqref{e.ctot} and \eqref{eq:sigtot} into the above.  
ATLAS and CMS also  made a number of exclusive studies where kinematic cuts  were employed to enhance the VBF contribution. 
In that case,  it is important to take into account the corresponding cut efficiencies $\eps_i$ for the different production channels.
For example for exclusive diphoton searches we have, 
\beq
R_{\gamma \gamma}^{\rm exc} = {\eps_{ggF} |\hat c_g|^2 \sigma_{ggF}^{\rm SM} / |\hat c_{g,\rm SM}|^2 +  \eps_{VBF}  |c_V|^2  \sigma_{VBF}^{\rm SM} +  \eps_{VH}  |c_V|^2 \sigma_{VH}^{\rm SM} 
\over   
\eps_{ggF}  \sigma_{ggF}^{\rm SM}  +  \eps_{VBF} \sigma_{VBF}^{\rm SM}  +  \eps_{VH}\sigma_{VH}} { {\rm Br}(h \to \gamma \gamma ) \over  {\rm Br}_{SM}(h \to \gamma \gamma) }.
\eeq  
The most prominent example is the dijet class of the CMS diphoton channel \cite{CMS_gaga}, where 2 forward jets with a large rapidity gap are required.  
In that case Monte Carlo simulations suggest $\eps_{ggF}/\eps_{VBF} \sim 0.03$, and $ \eps_{VH} / \eps_{VBF} \sim 0$.   Large systematic uncertainties are expected however.  
Another example is the ATLAS  fermiophobic Higgs search  \cite{ATLAS_fp},  where  $\eps_{ggF}/\eps_{VBF} \sim 0.3$. 
Thus, the ATLAS fermiophobic selection (much like the inclusive selection in the CMS fermiophobic search \cite{CMS_fp}, but unlike the CMS dijet tag class) is typically dominated by the ordinary ggF production mode, unless $c_g/c_V  \ll 1$. 

At the Tevatron the channel  most sensitive to a light Higgs signal is the $h \to b \bar b$ final state produced in association with a W/Z boson.
In this case the relevant rate is   
\beq
R_{b b}^{\rm Tev} \equiv {\sigma (p \bar p \to V h)  \over \sigma_{\rm SM} (p \bar p \to V h) }{ {\rm Br}(h \to b \bar b) \over {\rm Br}_{SM}(h \to b \bar b) } 
\simeq  \left|\frac{c_V c_b}{C_{tot}} \right |^2.  
\eeq 

Finally, it is interesting to consider the invisible Higgs rates at the LHC defined as  
\beq
R_{inv}^{ggF} \equiv {\sigma_{ggF} {\rm Br}(h \to \chi \bar \chi)  \over \sigma_{ggF}^{\rm SM} }, 
\quad 
R_{inv}^{VBF} \equiv {\sigma_{VBF} {\rm Br}(h \to \chi \bar \chi)  \over \sigma_{VBF}^{\rm SM} }.  
\eeq 
Currently, there is no official LHC limits on the invisible Higgs rate. 
Recasting the results of the LHC monojets searches one can arrive at the limits $R_{inv}^{ggF} < 1.9$, $R_{inv}^{VBF} < 4.3$ at 95\% CL \cite{Djouadi:2012zc}.
Combining ggF and VBF (assuming they come in the same proportions as in the SM), a somewhat stronger limit $R_{inv}< 1.3$ can be obtained. 
In any case, the currently available data can place a non-trivial direct constraint on the invisible Higgs branching fraction only in models where the Higgs production cross section is enhanced, for example in models with the 4th generation of chiral fermions where Higgs decays into 4th generation neutrinos~\cite{Kuflik:2012ai}. 
Alternatively, in a more model-dependent fashion, one can constrain the  invisible Higgs width  indirectly from the fact of observing the {\em visible} Higgs decays.
Assuming other Higgs couplings take the SM value,  ${\rm Br}(h \to \chi \bar \chi)$ larger than 50\% is disfavored \cite{Giardino:2012ww,Espinosa:2012vu}.

\section{Fits} 

\begin{figure}
\includegraphics[width=0.9\textwidth]{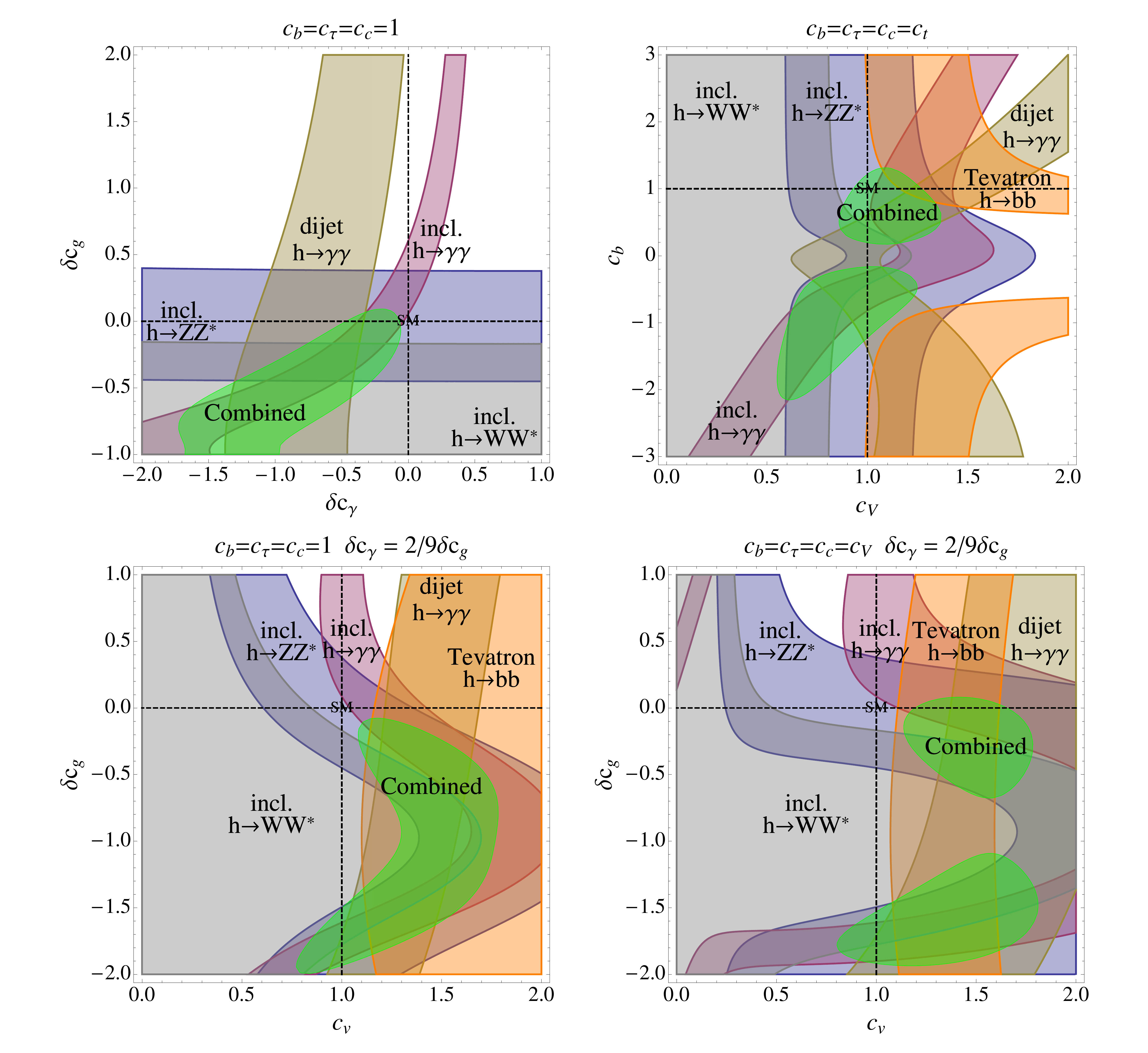}
\caption{
The allowed parameter space of the effective theory given in Eq.~\eqref{eq:1}  derived from the ATLAS, CMS  and Tevatron constraints for $m_h = 125$ GeV. 
We display the $1\sigma$ regions  allowed by the LHC inclusive $h \to \gamma\gamma$ channel (mauve), the LHC inclusive  $h \to ZZ^{*} \to 4l$ channel (indigo), 
the CMS dijet class of the $h \to \gamma\gamma$ channel (beige), the ATLAS inclusive $h \to WW^* \to 2l2\nu$ channel (light grey), and the Tevatron $h \to b \bar b$ W/Z boson associated channel (peach), 
 The lime green region is the one favored at Ê$90\%$ CL from the combination of these channels. The dashed lines show the SM values. 
 \label{fig:generalFit}}
\end{figure}

We are ready to  place constraints on the parameters of the effective theory. 
With enough data from the LHC one could in principle perform a full seven-parameter fit, however for the time being we pursue a simpler approach. 
Throughout we assume  $c_c = c_\tau = c_b$, and $c_{inv} = 0$, and study the LHC and Tevatron constraints on  the parameter space span by $\delta c_\gamma \equiv c_\gamma -  c_\gamma^{\rm SM}$, $\delta c_g \equiv c_g -c_{g,\rm SM}$, $c_V$, and $c_b$.  In this space, the best-fit point occurs for 
\beq
\delta c_\gamma \approx  -0.7, \quad \delta c_g \approx -0.5, \quad  c_V \approx 1.3, \quad  c_b \approx  -1.5,  
\eeq  
with  $\chi^2_{\rm min} \approx 0.9$.  
The SM  point $\delta c_\gamma = \delta c_g  = 0$,  $c_V = c_b = 1$ has  $\chi^2_{\rm SM} \approx  8.2$, which is disfavored at $88\%$ CL.

We also  study the best-fit regions in new physics models where only two of the above parameters can be freely varied, while the remaining ones are fixed. 
Sample results are displayed in Fig.~\ref{fig:generalFit}. 
In each plot the ``Combined'' region corresponds to $\Delta  \chi^2 < 4.6$, corresponding to the 90\% CL favored region in these models. 
The top left plot characterizes models in which loops containing beyond the SM fields contribute to the effective $h \, G_{\mu \nu}^a G_{\mu \nu}^a$ Êand Ê$h \, A_{\mu \nu} A_{\mu \nu}$ operators, while leaving the lower-dimension Higgs couplings in Eq.~\eqref{eq:1} unchanged relative to the SM prediction. 
Note that in these plots the Tevatron band are absent. That's because the Tevatron $b \bar b$ rate depends mostly on the parameters $c_b$ and $c_V$, and  very weakly on $c_g$ and $c_\gamma$.  Interestingly, in this section of the parameter space the Tevatron result is always {\em outside} the $1 \sigma$ band.   
In the remaining  plots  we fix $\delta c_\gamma = (2/9) \delta c_g$, which is the case in  {{top partner}} models where  only scalars and fermions with the same charge and color as the top quark contribute to these effective five-dimensional operators. 
The results are shown for three different sets of assumptions about the lower-dimension Higgs couplings that can be realized in concrete models addressing the Higgs naturalness problem. In particular, the assumptions in the  top-right plot are inspired by  composite Higgs models \cite{Giudice:2007fh}, where the couplings to the electroweak  gauge bosons and the couplings  all the SM fermions  are scaled by common factors, $c_V$ and $c_b$ respectively. 
The coupling to the top quark $c_t$ in the UV completion is also assumed to be rescaled by $c_b$, producing the corresponding shift of  $c_g$ and $c_\gamma$ in our effective theory. 
The interesting feature of this plot is the presence of two disconnected best-fit regions  \cite{Azatov:2012bz}. 
This reflects the degeneracy of the relevant Higgs rates in the $VV^*$ and $b \bar b$ channels under the reflection $c_b \to -c_b$,  which is broken only in the $\gamma \gamma$.   
Amusingly, a slightly better fit  is obtained in the $c_b < 0$ region, although it may be difficult to construct a microscopic model where such a possibility is realized naturally.    
It is worth noting that  the fermiophobic Higgs scenario, corresponding to $c_b = 0$ and  $c_V =1$, is  disfavored by the data (more generally, the fermiophobic line $c_b = 0$ is disfavored for any $c_V$). 
The two bottom plots  demonstrate  that the current data  show a preference for a slightly enhanced Higgs coupling to the electroweak gauge bosons, $c_V> 1$
and a slightly suppressed effective couplings to the gluons, $c_g  < 1$. 
This result is driven by the somewhat low  event rate (with respect to the SM) observed in the $WW^*$ and, to a lesser extent in the $ZZ^*$ channels (sensitive to the gluon fusion production), while  the data in the diphoton channel and in the Tevatron $b \bar b$ channel (sensitive to the Higgs coupling to W/Z),  are well above the SM expectations. 
Several well-studied models such as the MSSM or the minimal composite Higgs (and more generally, models with only SU(2) singlets and doublets in the Higgs sector), predict $c_V \leq 1$. 
If $c_V > 1$ is confirmed in the 8 TeV LHC run, it would point to a very specific direction for electroweak symmetry breaking \cite{Falkowski:2012vh}. 

To conclude, the LHC and Tevatron Higgs data have a great potential to test the consistency of the SM. 
With the limited statistics available, any conclusion about the Higgs couplings should be taken with a grain of salt.   Nonetheless, the analysis presented here demonstrates  the strength of constraining the effective Higgs Lagrangian as a mean to place bounds on new physics.  With more data we will soon learn whether the intriguing patterns currently visible shall disappear or rather  they are the  first signs of new physics.






\section*{Acknowledgements}

AF thanks the organizers of  the XXVI Rencontres de Physique de la Vall\'ee d'Aoste for the invitation and the view. 


\end{document}